\documentclass[twocolumn,floatfix,preprintnumbers,groupedaddress]{revtex4-1}

\usepackage[utf8]{inputenc}
\usepackage[colorlinks=true,citecolor=blue,linkcolor=blue]{hyperref}
\usepackage[normalem]{ulem}
\usepackage{amsmath,amssymb,mathtools}
\usepackage{epsfig}
\usepackage{graphicx}               
\usepackage{url}
\usepackage{color,colortbl}
\usepackage{slashed}
\usepackage{cancel}
\usepackage{multirow}
\usepackage{placeins}
\usepackage[dvipsnames]{xcolor}
\usepackage{epstopdf}
\usepackage{soul}
\usepackage{tikz}
\usepackage{tikz-cd}
\usepackage{tikz-feynman}
\usepackage[capitalise, english]{cleveref}
\usepackage{siunitx}
\usepackage{xspace}
\usetikzlibrary{trees}
\usetikzlibrary{decorations.pathmorphing}
\usetikzlibrary{decorations.markings}

\definecolor{red}{rgb}{0.6,.0706,.1373}
\definecolor{blue}{rgb}{0,0.396,0.741}
\newcommand\myshade{80}
\colorlet{mylinkcolor}{magenta}
\colorlet{mycitecolor}{magenta}
\colorlet{myurlcolor}{blue}

\hypersetup{
  linkcolor  = mylinkcolor!\myshade!black,
  citecolor  = mycitecolor,
  urlcolor   = myurlcolor,
  colorlinks = true
}

\allowdisplaybreaks

\newcommand{\Z}{\mathbb{Z}}
\newcommand{\R}{\mathbb{R}}
\newcommand{\be}{\begin{equation}}
\newcommand{\ee}{\end{equation}}
\newcommand{\bea}{\begin{eqnarray}}
\newcommand{\eea}{\end{eqnarray}}

\def\L{\mathcal{L}}

\def\beq#1\eeq{\begin{align}#1\end{align}}

\makeatletter
\providecommand*{\diff}%
  {\@ifnextchar^{\DIfF}{\DIfF^{}}}
\def\DIfF^#1{%
  \mathop{\mathrm{\mathstrut d}}%
    \nolimits^{#1}\gobblespace}
\def\gobblespace{%
  \futurelet\diffarg\opspace}
\def\opspace{%
  \let\DiffSpace\!%
  \ifx\diffarg(%
    \let\DiffSpace\relax
  \else
    \ifx\diffarg[%
      \let\DiffSpace\relax
    \else
        \ifx\diffarg\{%
        \let\DiffSpace\relax
      \fi\fi\fi\DiffSpace}


\begin{document}

\title{A Composite Theory of Higgs and Flavour}

\hfill~CERN-TH-2025-267,\, ZU-TH~88/25

\medskip

\author{Joe Davighi}
\email{joseph.davighi@cern.ch}
\affiliation{Theoretical Physics Department, CERN, 1211 Geneva 23, Switzerland}
\author{Gino Isidori}
\email{isidori@physik.uzh.ch}
\affiliation{Physik-Institut, Universit\"at Z\"urich, CH 8057 Z\"urich, Switzerland}


\preprint{}

\begin{abstract}
We introduce a composite Higgs model in which a flavour deconstructed gauge group is embedded in the strong sector. The pattern of global symmetry breaking yields, as pseudo-Nambu–Goldstone (pNGB) bosons, both a Standard Model (SM)-like Higgs and the link field whose vacuum expectation value breaks the flavour non-universal gauge group down to the SM. Inevitably, a third pNGB appears, transforming as a second Higgs doublet coupled only to the light generations. This heavy Higgs naturally mediates suppressed light Yukawa couplings, providing a solution to the flavour puzzle. At the same time, new physics contributions to flavour violating observables are CKM- and chirally suppressed, while electric dipole moment bounds are evaded through an automatic mass alignment in the light fermion sector. The result is a natural framework for addressing the origin of both Higgs and flavour hierarchies, with reduced Higgs mass tuning and minimised impact in flavour observables, that is best tested by high-$p_T$ and precision electroweak measurements.

\end{abstract}

\maketitle

\section{Introduction} \label{sec:intro}

Composite Higgs models provide an attractive solution to the electroweak hierarchy problem by identifying the Higgs field of the 
Standard Model (SM) with a pseudo-Nambu--Goldstone  boson  (pNGB) of a new strongly coupled sector~\cite{Agashe:2004rs,Contino:2003ve,Panico:2015jxa}. Precision electroweak measurements and direct collider searches require the new degrees of freedom to lie above a few TeV~\cite{ParticleDataGroup:2024cfk}, implying the presence of an unavoidable little hierarchy between the Higgs mass and the scale $f$ that characterises the strong dynamics. Understanding how this residual tuning can be minimised while remaining compatible with 
the tight constraints from flavour- and CP-violating observables is one of the central challenges facing realistic  constructions.

Flavour physics plays a crucial role in this context. One of the original motivations for compositeness was the possibility of addressing the flavour puzzle through partial compositeness~\cite{Kaplan:1991dc}, whereby the composite dynamics is flavour anarchic and the observed fermion mass hierarchies are the consequence of a suitably chosen linear mixing between elementary and composite fermions. However, in its simplest realisation, this mechanism is severely constrained by flavour- and CP-violating observables. In particular, electric dipole moments (EDMs) and light-flavour mixing generically push the scale of the strong dynamics well above 100~TeV, far beyond what is compatible with a natural Higgs sector.

It has long been known that imposing global flavour symmetries can substantially alleviate these constraints. In composite Higgs models with a $U(3)^5$ flavour symmetry the scale of compositeness can be lowered to $M_\ast \sim 7$~TeV, while $U(2)^n$ symmetries allow values as low as $M_\ast \sim 2$~TeV, minimizing the little hierarchy~\cite{Glioti:2024hye} (see also Ref.~\cite{Agashe:2025tge}). These constructions, however, typically postulate the presence of such global flavour symmetries without providing a dynamical explanation for their origin.

An appealing possibility is that these flavour symmetries emerge as accidental low-energy properties from an enlargement of the gauge structure. This idea can be realised, for example, through the 
flavour deconstruction hypothesis (see~\cite{Davighi:2023iks,Barbieri:2023qpf} and references therein), assuming the SM gauge group is the flavour universal subgroup of a 
manifestly flavour non-universal gauge symmetry in the 
ultraviolet (UV). For instance, SM hypercharge can arise as the diagonal subgroup of $U(1)_{Y}^{[12]} \times U(1)_{Y}^{[3]}$~\cite{Davighi:2023evx,FernandezNavarro:2023rhv}. This hypothesis can simultaneously explain the observed hierarchies in the Yukawa couplings and the smallness of the light-heavy mixing in the 
Cabibbo-Kobayashi-Maskawa (CKM) mixing matrix, while enforcing $U(2)$-like flavour symmetries for the light generations which prevent large non-SM effects in low-energy observables.

In this work, we explore how flavour deconstruction can be embedded consistently within non-minimal composite Higgs models (for previous attempts, see~\cite{Fuentes-Martin:2020bnh,Fuentes-Martin:2022xnb,Chung:2021fpc,Chung:2023iwj,Chung:2023gcm,Covone:2024elw,Lizana:2024jby}). We enlarge the global symmetry of the strong sector so that the deconstructed gauge symmetry appears as a subgroup, with the corresponding link field arising naturally as a pNGB. The deconstruction in flavour space involves only the hypercharge sector of the SM, while $SU(2)_L$ and $SU(3)_c$ are kept flavour universal~\cite{Davighi:2023evx,Barbieri:2023qpf,FernandezNavarro:2023rhv}. Remarkably, this setup automatically provides all the ingredients required for a realistic Yukawa sector: in addition to the SM–like Higgs, a second, heavier Higgs doublet emerges as a pNGB and mediates the Yukawa couplings of the light generations.

The resulting structure has important phenomenological consequences in both Higgs and flavour sectors. On the flavour side, non-SM contributions to light–heavy  transitions, such as $B_s \to \mu^+ \mu^-$,  are strongly suppressed by the flavour universality of left-handed interactions (as in the SM). Flavour-changing couplings involving light fermions are mediated solely through the heavy Higgs doublet and are both CKM and light-Yukawa suppressed. Since all $U(2)$-breaking effects are aligned with the SM Yukawa matrices, the model introduces no new sources of CP violation, rendering electric dipole moment constraints automatically harmless. As a result, the scale of compositeness can naturally be brought close to the TeV. 
On the Higgs side, as already pointed out in~\cite{Covone:2024elw}, the extended gauge sector provides a more natural framework to cancel gauge and fermion contributions in the effective potential of the SM-like Higgs field. In this context, the smallness of the Higgs mass is the result of an approximate accidental cancellation in a $3\times 3$ pNGB mass matrix, where the two other eigenvalues (the link field and the heavy Higgs) 
have natural size of $O(f^2)$.

\section{The Model}

\subsection{Left-universal gauge structure}

In~\cite{Davighi:2023iks,Barbieri:2023qpf} it has been argued that, within the framework of flavour deconstruction, a model with deconstructed $SU(2)_R$ and $U(1)_{B-L}$ symmetry but flavour universal $SU(3)_c\times SU(2)_L$ offers the best starting point for explaining both the SM and BSM flavour puzzles. It predicts Yukawa textures of the form
\begin{equation} \label{eq:yuk}
    Y \sim \begin{pmatrix}
         \epsilon_R & \epsilon_{\Omega} \\
        \epsilon_R \epsilon_{\Omega} & 1
    \end{pmatrix}
\end{equation}
where $\epsilon_R$ ($\epsilon_{\Omega}$)
generically 
denotes the norm of the  matrix (vector) 
acting on the $2\times 2$ subspace of the light families.
This structure has the following desirable features: (i) independent spurions sourcing $|V_{cb}|\ll 1$ (from $B-L$ deconstruction) and $y_2\ll y_3$ hierarchies (from $SU(2)_R$ deconstruction); (ii) doubly-suppressed right-handed mixing, which helps soften flavour bounds in the presence of other new physics particles associated {\em e.g.} with the composite sector, (iii) new gauge bosons with reasonably weak experimental constraints compared to deconstructing $SU(2)_L$~\cite{Davighi:2023evx,Davighi:2023xqn}, viable at a lower scale.

We take this starting point seriously, and show how it can be combined with the hypothesis of Higgs compositeness. From the point of view of the weak scale, it makes sense to treat $SU(2)_R$ and $U(1)_{B-L}$ differently, given only the $SU(2)_R$ deconstruction has relevant implications for the stability of the electroweak scale. The
$SU(2)_R$ gauge bosons ($V_R^\mu$) generate 
contributions to the Higgs mass already 
at the one-loop
level; moreover, if the Higgs is to be embedded as a pNGB, we need $SU(2)_R \subset G_{\mathrm{glob}}$. So, it is desirable to include the full deconstructed $SU(2)_R$ dynamics inside the global symmetry of the strong dynamics.
On the other hand, the deconstruction of $U(1)_{B-L}$ can be disentangled from the composite sector without much consequence for naturalness; it can be associated to a symmetry breaking occurring
at a UV scale from which the pNGB Higgs would be safely protected.

\subsection{Symmetry breaking pattern}

With this rationale, we consider the following scenario:
\begin{itemize}
    \item We suppose the \textbf{UV} dynamics has a
    global symmetry 
    \begin{equation}
        G=Sp(6) \times U(1)_{B-L}^{[12]} \times U(1)_{B-L}^{[3]},
    \end{equation}
    with $SU(2)_L \times SU(2)_R^{[12]}\times SU(2)_R^{[3]}$ maximally embedded in the $Sp(6)$ factor. 
    We envisage there is no fundamental scalar in the theory, only fermions and gauge fields. 
    Within $G$, we suppose that only the subgroup
    \begin{equation} \label{eq:gauging}
        K=SU(2)_L \times U(1)_Y^{[12]}\times U(1)_R^{[3]} \times U(1)_{B-L}^{[3]}
    \end{equation}
    is gauged.
    \item At some \textbf{high scale} $\gg$ TeV, we assume the following symmetry 
    breaking occurs
    \begin{equation} \label{eq:BL_break}
        \langle \Omega \rangle : U(1)_Y^{[12]}\times U(1)_{B-L}^{[3]} \longrightarrow U(1)_X,
    \end{equation} 
    where $X = Y_{12}+(B-L)_3$. We remain agnostic about the origin of $\langle \Omega \rangle$; 
    it could for instance originate from some fermion condensate.
    Below the scale $\langle \Omega \rangle$ the three generations of left-handed (LH) fermions look the same from the gauge point of view. However, there remains an accidental symmetry $U(1)_{B-L}^{[12]}\times U(1)_{B-L}^{[3]}$, broken by the spurion
    \begin{equation}
    \epsilon_\Omega \equiv \langle \Omega \rangle /M_\Psi \sim |V_{cb}| \approx 0.04\,,
    \label{eq:e_omega}
    \end{equation}
    where $\Psi$ is a vector-like fermion in the UV.
    \item Running down towards the \textbf{TeV scale}, we suppose there is a global symmetry breaking transition due to some strong dynamics,
    \begin{equation}
    \frac{G}{H} = \frac{Sp(6)}{SU(2)_L \times SU(2)_{R,12} \times SU(2)_{R,3}}\, ,
    \end{equation}
    occurring at a scale $4 \pi f$, desiring $f$ to be as close to the TeV as is viable. (In this work, we do not put forward an explicit dynamical origin for this global symmetry breaking pattern.)
\end{itemize}
    
The global breaking pattern at $4\pi f$ delivers pNGBs in the following representations of $H$:
\begin{align} \label{eq:pNGBs}
    H_{3} &\sim ({\bf 2,1,2 }) \qquad \text{
    (SM-like Higgs)}\,, \\
    H_{12} &\sim ({\bf 2,2,1 }) \qquad \text{ (Heavy Higgs)}\,, \nonumber \\
    \Sigma &\sim ({\bf 1,2,2 }) \qquad \text{ (Link field)\,.} \nonumber
\end{align}
The field $\Sigma$ has the quantum numbers of the $SU(2)_R$ link field, essential to flavour deconstruction. Due to explicit $Sp(6)$-breaking, a potential is generated for all the pNGBs, with loops roughly cut-off by $4\pi f$ (but always suppressed by spurions). For the quadratic terms, it is just as possible to get a positive or negative sign coefficient, given we have sizable contributions from both gauge and fermion loops. 
We consider a scenario (see below) in which $\Sigma$ gets a vev $\langle \Sigma \rangle \sim f$ to trigger a breaking of global symmetries $SU(2)_{R}^{[12]}\times SU(2)_{R}^{[3]} \to SU(2)_R$. Taking the intersection with $K$, this implies
\begin{equation}
    \langle \Sigma \rangle\, :\, U(1)_{X}\times U(1)_{R}^{[3]} \to U(1)_Y 
\end{equation}
landing on the SM gauge symmetry.
This completes the overall symmetry breaking scheme of our model, above the electroweak scale. 

\subsection{The one-loop pNGB potential}

Our setup predicts several states near the scale $f\sim$ TeV.
In addition to the $V_R^\mu$ gauge bosons, which get mass $M_R \approx g_{R_3} \langle \Sigma \rangle$ from the longitudinal mode of $\Sigma$, there is a second Higgs doublet -- a pNGB multiplet transforming in the same gauge representation as the SM Higgs.
Above the scale $\langle \Sigma \rangle$, the gauge symmetry dictates that $H_3$ couples only to third generation fermions while $H_{12}$ only to the light generations.

As mentioned, all three pNGB multiplets get their mass at one-loop, due to explicit breaking of $Sp(6)$ global symmetry by gauging and interactions with fermions.
To match observations, the interesting r\'egime is that where $H_{12}$ does not get a vev and has an untuned mass of order $f\sim \text{TeV}$, while the $H_3$ must get a tuned vev of order $v \sim (0.1 f)$, as per the little hierarchy problem:
\begin{equation} \label{eq:quads}
    M_{3}^2 \sim -(0.1 f)^2, \quad M_{12}^2 \sim +f^2, \quad M_{\Sigma}^2 \sim - f^2
\end{equation}
A positive and untuned $M_{12}^2$ is in fact `predicted' in our model setup: as we show in \S \ref{sec:model_yuk}, $H_{12}$ has only small couplings to fermions, hence $M_{12}^2$ is dominated by the (necessarily positive) gauge contribution. 
The tuning in $M_3^2$ is between the fermion and gauge contributions, as is typical and necessary for the pNGB Higgs. Assuming the top Yukawa arises via partial compositeness~\cite{Kaplan:1991dc}, and that the linear mixing responsible satisfies a parity symmetry that renders the fermion loop finite~\cite{Csaki:2017cep}, we expect
\begin{equation} \label{eq:Hmass}
    M_3^2 \approx \frac{1}{4\pi^2}\left(-N_c y_t^2 M_{T}^2 + \frac{9}{8}(g_L^2+g_{R_3}^2) M_{\rho}^2 \right)
\end{equation}
with $M_T$ the mass of the top partner and $M_\rho$ the lightest spin-1 resonance that cuts off the gauge loops.
We further discuss the quadratic potential terms, and the sort of dynamical hypotheses that would lead to~\eqref{eq:quads}, in App.~\ref{app:quads}. 

A feature that arises naturally in our setup is that the three different scalar multiplets talk via a cubic interaction generated in the one-loop potential. While this is a general feature,  
the strength of the corresponding interaction 
depends on UV details. As shown in App~\ref{app:quads}, adding a vector-like fermion, $\chi$, charged under $U(1)_Y^{[12]}$ and $U(1)_{B-L}^{[3]}$, leads to
\begin{equation} \label{eq:cubic}
    V \supset -\frac{ N_c}{4\pi^2} \frac{M_{T}^2}{M_\chi} \, \mathrm{Tr}\, H_3 \Sigma H_{12}^\dagger\,
\end{equation}
up to order-one couplings. 
This cubic interaction mediates the dominant interaction between $H_3$ and the light fermion sector. As we discuss next, for 
$M_\chi \gtrsim M_T$ it has the correct size to 
generate the desired flavour structure.

\subsection{Yukawa hierarchies and \boldmath{$U(2)$} protection} \label{sec:model_yuk}

The flavour-deconstructed gauge symmetry, together with the coset construction, allows us to fix the most general Lagrangian that is bilinear in SM fermions in terms of form factors --  irrespective of our dynamical assumptions on how these interactions are generated (which we set out in due course). 
Those form factors that flip the fermion chirality give rise to Yukawa couplings, {\em e.g.}
\begin{align} \label{eq:chiral_flip}
    \L \supset\,  &\overline{q}_L^i \,\, Y_{i3}^u(p^2) f \sin\left(\frac{h_{3}}{f}\right) \cos\left(\frac{h_{3}}{f}\right)\, t_R + (u,t \leftrightarrow d,b) \nonumber \\
    + &\overline{q}_L^i \,\, Y_{i2}^u(p^2) f \sin\left(\frac{h_{12}}{f}\right) \cos\left(\frac{h_{12}}{f}\right)\, c_R\, + \dots
\end{align}
The form of the trigonometric functions
is fixed by the coset construction; we use a shorthand $\sin(h/f) = \sin(|H|/f)H/|H|$, with $|H|=\sqrt{\mathrm{Det}(H)}$ and $H=H^4 + \sum_{a=1}^3\sigma^a H^a$  
{\em etc}. To leading order, the different pNGB multiplets do not mix in this fermion bilinear term.

The high-scale breaking of the non-universal $B-L$ implies that $Y_{i3}^{u,d}(0) \sim (\epsilon_\Omega x^T_{u,d},\,\, y_{t,b})$, 
where $\epsilon_\Omega$ is the spurion  in \eqref{eq:e_omega}, and 
 $x_{u,d}$ are two-component vectors. 
A similar 2+1 structure holds for the $h_{12}$ couplings; however, in this case it is the entry involving the third-generation that is suppressed by $\epsilon_\Omega$.

\medskip

\paragraph{Third generation Yukawas.}
Given the largeness of the top Yukawa coupling, it is phenomenologically safest to suppose $y_t$ arises from partial compositeness (PC)~\cite{Kaplan:1991dc,Panico:2015jxa}. 
This implies
\begin{equation} \label{eq:Y3}
    Y_{i3}^{u,d} \sim \frac{f}{M_\ast} \lambda^L_i \lambda_{t, b}^{R\ast} \sim \frac{1}{g_\ast} \lambda_i^L(\Lambda) \lambda_{t,b}^{R\ast}(\Lambda) \left(\frac{M_\ast}{\Lambda}\right)^{\gamma_L - \gamma_R}
\end{equation}
where $\gamma_{L,R}$ are the anomalous dimensions for the linear mixing operators, which we expect to cluster around zero. A small difference in anomalous dimensions could easily generate the large splitting within the third generation masses, $y_b \ll y_t$.
The scale $M_\ast$ in~\eqref{eq:Y3} might be replaced by a particular low-lying resonance, {\em e.g.}~the top partner, if we assume that dominates the two point function. 

\medskip

\paragraph{Light Yukawas.}
For phenomenological reasons that should become clear, we suppose the interactions indicated by the second line of~\eqref{eq:chiral_flip}, {\em i.e.} the couplings of the $H_{12}$ heavy multiplet to the light generations, do {\em not} arise from PC but rather via a `direct' coupling of a scalar composite operator to the fermion bilinear, as 
proposed in~\cite{Panico:2016ull}.
That is, above the scale of the global $Sp(6)$ breaking, we suppose there are couplings of the form
\begin{equation} \label{eq:protoY}
    \mathcal{L}\supset \underbrace{\alpha_{ij}\left(\frac{M_\ast}{\Lambda}\right)^{\gamma_{LR}}}_{\hat{y}_{ij} } \overline{\psi}_{L}^{i}\mathcal{O}_{12}\psi_R^j \quad i,j\in\{1,2\}    \, .
\end{equation}
where $\mathcal{O}_{12}$ is a scalar operator with the same quantum numbers as the $H_{12}$ field, that will interpolate to this pNGB at low-energy.
These couplings are gauge-invariant and moreover unsuppressed by the $B-L$ breaking spurion. (For $i=3$, an insertion of $\epsilon_{\Omega}$ is required as mentioned.) 

In~\eqref{eq:protoY}, $\gamma_{LR}$ denotes the quantum anomalous dimension of the operator there written. We expect $\dim(\mathcal{O}_{12}) \gtrsim 2$ and 
thus $\gamma_{LR} \gtrsim 1$~\cite{Rattazzi:2008pe,Rychkov:2009ij}, in other words these direct bilinears behave like dimension-5 (at least) effective operators in the low-energy theory. 
This endows the proto-Yukawa couplings $\hat{y}_{ij}$ in~\eqref{eq:protoY} with some scale suppression already. 
For instance, we might suppose the next scale $\Lambda$ is an order of magnitude above $M_\ast$, which would mean $|\hat{y}_{ij}| \lesssim 0.1$.
Since the light Yukawa couplings are small, this is not a problem but rather a feature -- to be contrasted with the need for partial compositeness in order to get a large enough top Yukawa (for a large enough scale $\Lambda$).

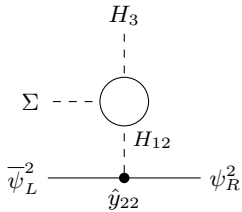
\begin{figure}[t]
    \begin{tikzpicture}[baseline=-6.0ex]
    \begin{feynman}
        \vertex (a) {$H_{3}$};
        \vertex [below=0.45in of a] (b);
        \vertex [left=0.4in of b] (c) {$\Sigma$};
        \vertex [below=0.4in of b] (d);
        \vertex [below=0.01in of d] (d1) {$\hat{y}_{22}$};
        \vertex [left=0.4in of d] (e) {$\overline{\psi}_L^2$};
        \vertex [right=0.4in of d] (f) {$\psi_R^2$} ;
        \diagram* {
        (a) -- [scalar] (b) -- [scalar,edge label=\footnotesize{$H_{12}$}] (d),
        (b) -- [scalar] (c),
        (f) -- (d) -- (e)
        };
        \node at (d) [circle,fill,inner sep=1.4pt]{};        
        \node at (b) [circle,fill,white,inner sep=6.5pt]{};
        \node at (b) [empty dot,inner sep=6.5pt]{};                       
    \end{feynman}
    \end{tikzpicture}
\caption{\label{fig:one}
Effective Yukawa interaction for the light fermions.}
\end{figure}

Now we are ready to understand the origin of the hierarchical Yukawa structure~\eqref{eq:yuk} in the model.
Together with the cubic mixing~\eqref{eq:cubic}, the interactions~\eqref{eq:protoY} imply that the SM-like Higgs ($H_3$) will talk to the light fermions, via the exchange of $H_{12}$ and an insertion of $\langle \Sigma \rangle$, as shown in Fig.~\ref{fig:one}.
Upon integrating out the heavy $H_{12}$, we get 
\begin{align}
    \mathcal{L}_{\mathrm{EFT}} &\supset y_{ij}\, {\bar \psi}_{L}^{i}H_{3}\psi_R^j\,, 
    \qquad
    y_{ij} = \kappa\, \hat{y}_{ij}\,,
    \nonumber \\
    \kappa & \approx \frac{N_c}{4\pi^2} 
    \frac{ v_\Sigma M_T^2 }{ M_\chi M^2_{12}} 
    =O(10^{-1})\,,
\end{align}
for $i,j\in\{1,2\}$.
The power suppression of $\hat{y}_{ij}$ as explained above, 
together with the loop-suppression of the cubic vertex in the potential, leads to the small spurion 
\begin{equation}
    \epsilon_R \sim \kappa  \times | {\rm max} (\hat{y}_{ij})|
    \sim 10^{-2} \quad
\end{equation}
that sets the overall size of the light-fermion 2-by-2 sub-block of each Yukawa matrix. We
thus recover the texture in Eq.~\eqref{eq:yuk}.

\medskip

\paragraph{$U(2)$ protection.}

Our hypothesis that the light Yukawa couplings arise from direct fermion bilinears, rather than from partial compositeness, is not only instrumental in reproducing the observed hierarchies of the Yukawa matrix. Equally important, this mechanism ensures a minimal breaking of the $U(2)^5$ flavour symmetry acting on the light generations. As defined in~\cite{Barbieri:2011ci}, this minimal breaking is characterised by a spurion
bidoublet associated to the light Yukawa eigenvalues ($\epsilon_R$), and 
a spurion transforming as a doublet under the $U(2)$ symmetry acting on left-handed fermions ($\epsilon_\Omega$). This is nothing but the structure we have in Eq.~(\ref{eq:yuk}). 

This breaking structure provides a Minimal Flavour Violating~\cite{DAmbrosio:2002vsn} (MFV)-like protection for flavour-changing neutral current (FCNC) processes. In particular, the uniqueness of the spurion controlling the light flavour sector implies that the $2\times2$ sector of the light families is diagonalised in the same basis both for the Yukawa couplings and for FCNC amplitudes induced by higher-dimensional operators. As a result, the only non-trivial contributions to FCNCs arise from heavy--light mixing, which is suppressed in a CKM-like manner. Moreover, flavour-violating amplitudes involving light right-handed fields are necessarily chirally suppressed ({\em i.e.}~proportional to the corresponding light Yuakwa couplings). 

\begin{figure}[t]
    \begin{tikzpicture}[baseline=-6.0ex]
    \begin{feynman}
        \vertex (a) {$s_L$};
        \vertex [right = 0.3in of a] (a1);
        \vertex [below=0.3in of a1] (b);
        \vertex [below=0.6in of a] (c) {$d_R$};
        \vertex [right = 0.4in of b] (d);
        \vertex [right = 0.4in of d] (d1);
        \vertex [right = 1.0in of a] (e) {$s_R$};
        \vertex [right=1.0in of c] (f) {$d_L$};
        \diagram* {
        (a)  -- [fermion] (b),
        (b) -- [fermion] (c),
        (b) -- [scalar, edge label=$H_{12}$] (d),
        (e) -- [fermion] (d),
        (d) -- [fermion] (f),
        };
        \node at (d) [circle,fill,inner sep=1.4pt]{};  
        \node at (b) [circle,fill,inner sep=1.4pt]{};                   
    \end{feynman}
    \end{tikzpicture}
\caption{\label{fig:DS2} $H_{12}$-mediated contribution to $\Delta S=2$ amplitudes.}
\end{figure}
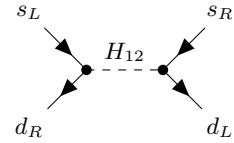

To better illustrate these protection mechanisms, we analyse  
more explicitly the case of $\Delta S = 2$ amplitudes. Here a potentially large tree-level contribution is induced  by the exchange of $H_{12}$ (see Fig.~\ref{fig:DS2}).
 However, taking into account the flavour structure of the 
 $H_{12}$ couplings in the mass-eigenstate basis (App.~\ref{app:H12}),
we get 
\begin{equation}
    \mathcal{A}_{\Delta F=2} [H_{12}] \;\sim\; \frac{1}{\kappa^2}\, y_s y_d \,(V^*_{td} V_{ts})^2 \, .
\end{equation}
Despite the  $\kappa^{-2} \sim 10^2$ enhancement, this contribution is safely suppressed by both the smallness of the light Yukawa couplings and by the CKM factors.

In principle, additional tree-level contributions to FCNCs are induced by the $Z^\prime$ exchange; however, the flavour universality of the $SU(2)_L$ interaction implies the FCNC couplings of the $Z^\prime$ are negligible (see App.~\ref{app:Zprime}). Finally, additional contributions are generated by the composite sector: these are characterised by 
an effective scale $M_\rho \gtrsim$~8~TeV. Given the spurion structure, the largest contribution occurs when only LH fields are involved, leading to the SM-like suppression factor 
$V^*_{3i} V_{3j}$ for each FCNC bilinear. Combined with the heavy  effective scale, also in this case the effect is below current bounds.

The uniqueness of the spurion controlling the light flavour sector also ensures a strong suppression of the EDMs of 
light quarks and leptons. Firstly, in the limit where mixing with the third generation is neglected, the $2\times2$ light-flavour sector  can be made real (no physical CP-violating phases survive after field redefinitions). Secondly, the effective operators associated to EDMs necessarily involve light right-handed fields and hence are chirally suppressed.

\section{Phenomenology}

This model predicts many phenomenological signatures characteristic to pNGB Higgs theories, equipped with minimally-broken $U(2)$ flavour symmetries, but with the addition of the heavy gauge boson $V_R^\mu$. 
The need to reproduce a realistic Higgs potential also suggests there should be a relatively light top partner, but the ingredient of flavour deconstruction allows this to be rather heavier than usual~\cite{Covone:2024elw}, giving better compatibility with LHC direct searches.

Concerning the impact of composite dynamics
on electroweak precision observables (EWPOs) and precision flavour observables, the benchmark identified in~\cite{Covone:2024elw} applies just as well to the present case:
\begin{equation}
    f\approx 1.5\mathrm{~TeV}, \quad M_T \approx 2\mathrm{~TeV}, \quad g_\rho \approx 5\,,
\end{equation}
with $g_\rho$ safely lower than $4\pi$ 
allowing a power counting related
to the loop expansion.
This benchmark corresponds to a~3\% tuning in the Higgs potential, per-mille contributions to EWPOs, and $M_\rho \sim 8$ TeV.

\begin{figure}[t]
    \centering
    \includegraphics[width=0.75\linewidth]{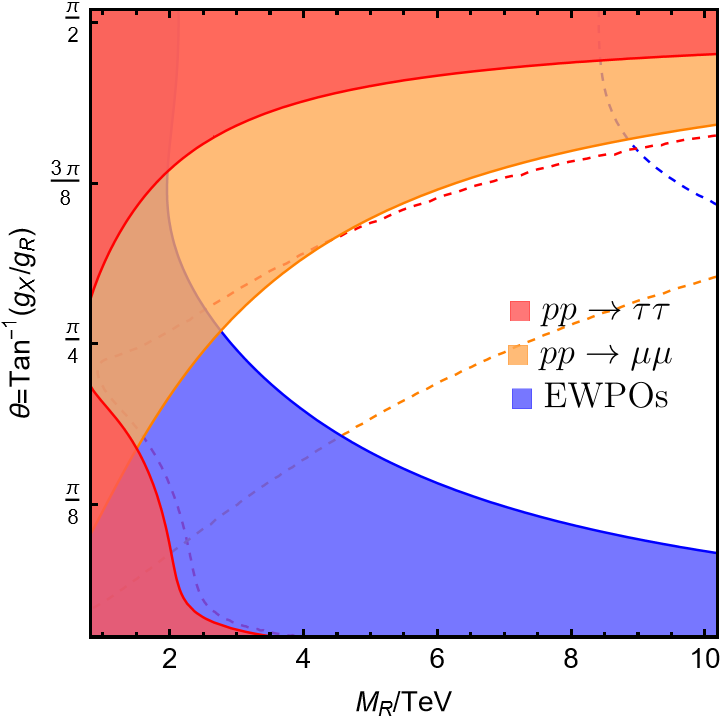}
    \caption{Key constraints on the heavy neutral gauge boson predicted by the flavour deconstructed sector of the model. The dashed lines correspond to projections, assuming SM-like measurements, from HL-LHC (for $pp\to \ell\ell$) and FCC-ee (for EWPOs).
    }
    \label{fig:Zp_bounds}
\end{figure}

The gauge sector is extended by a single heavy $Z^\prime$ boson with flavour-dependent couplings and mass $M_R$. As shown in App.~\ref{app:Zprime}, the flavour bounds on this $Z^\prime$ are very weak thanks to (i) its couplings to LH fermions being flavour-universal, and (ii) the doubly-suppressed RH mixing. 
The leading bounds are thus high-$p_T$  searches at the LHC, especially in $pp \to \mu^+\mu^-$~\cite{CMS:2021ctt} and $pp \to \tau^+ \tau^-$~\cite{ATLAS:2020zms}, plus EWPO measurements.
We compute the LHC bounds using the \texttt{HighPT} software~\cite{Allwicher:2022mcg,Allwicher:2022gkm}, by matching $Z^\prime$--mediated amplitudes onto SMEFT local operators. 
The EWPO constraint is derived using the likelihood from~\cite{Breso-Pla:2021qoe}, which we have updated to include the most recent $m_W$ combination~\cite{CMS:2024lrd,Reina:2025suh}. 
More details on the SMEFT operator matching that goes into these fits is recorded in App.~\ref{app:Zprime}. 

We summarise the bounds from current data in Fig.~\ref{fig:Zp_bounds}. The lightest viable mass is around
\begin{equation}
    M_{R} \gtrsim 3 \text{~TeV,} \qquad \text{for~~}     g_X \approx g_R \approx \sqrt{2} g_Y\, .
\end{equation}
At this point in parameter space, the constraint from $pp\to \mu\mu$ (which excludes large $g_X$) crosses the constraint from EWPOs (which excludes large $g_R$). 
As a result of its high precision, the $W$ mass measurement plays a leading role in shaping the EWPO bound in the region $\theta < \pi/4$. The modified $Z\tau_R\tau_R$ and $Zb_R b_R$ couplings also play a significant role in this region. For $\theta > \pi/4$ there are significant shifts in $Z$ couplings to light fermions, but in this region the LHC constraints have already taken over.

Looking to the future, these leading constraints on our model stand to improve significantly with proposed future experiments. We indicate the expected reach of HL-LHC with $3\, \text{ab}^{-1}$ data for the high-$p_T$ bounds (by performing a na\"ive rescaling of statistics with luminosity), and the expected sensitivity of FCC-ee for the EWPOs~\cite{Selvaggi:2025kmd}. In both cases, we obtain projections by assuming SM-like central values will be measured.
The future reach is extremely promising, well exceeding 10 TeV in $M_R$, thanks to the excellent complementarity between the high-$p_T$ and EWPO probes.

\bigskip

\section{Summary and Outlook}

In this paper we proposed that the flavour puzzle and the hierarchy problem can be solved together near the TeV scale, by embedding a flavour deconstructed $SU(2)_R$ symmetry inside a strong sector of which the Higgs is a composite pNGB, while keeping $SU(2)_L$ flavour-universal (as in the SM). 
In order to realise both the SM Higgs ($H_3$) and the link field ($\Sigma$) as pNGBs, where the latter is needed to break the non-universal gauge interaction to the SM, we are led to a breaking pattern that necessarily produces also a second pNGB Higgs ($H_{12}$) coupled to the light generation right-handed fermions.

The presence of $H_{12}$ is central to the phenomenological viability of the model. Through the cubic interaction, $\text{Tr}(H_3 \Sigma H_{12}^\dagger)$, its fermionic couplings are transmitted to the SM-like Higgs, but are naturally suppressed by the loop-induced nature of this operator. Assuming that the $H_{12}$ couplings originate from fermion bilinears of the form $\bar{f}\mathcal{O}_{12}f$, expected to arise from operators of dimension five or higher, allows us to achieve two key results: (i) realistic Yukawa hierarchies without introducing additional vector-like fermions or parameter tuning; (ii) a minimal breaking of the $U(2)^5$ flavour symmetry, which is known to ensure a phenomenologically safe flavour structure at the TeV scale. 
In particular, FCNC observables are CKM-suppressed (as in the SM) and, when involving right-handed fields, also chirally-suppressed. The same structure also implies the contributions to the EDMs of the light families are well beyond current bounds. Finally, the fact that the $Z^\prime$  boson arising 
from  deconstruction couples universally to left-handed fields implies  its contributions to flavour-changing  observables are also irrelevant for $M_{R} \sim$ TeV. 

As a result, the model can best be tested through high-$p_T$ and electroweak precision measurements. In addition to the standard contributions from the strong dynamics and composite resonances, here we have the additional contributions from the heavy gauge boson, currently viable for $M_R \gtrsim 3$ TeV. Due to enhanced gauge contributions to the Higgs mass, the top partner can be 
slightly heavier than in flavour universal scenarios, giving better compatibility with current null results from direct searches for top partners.

The model we have presented can be viewed as a TeV scale extension of the minimal flavour deconstructed gauge frameworks proposed in~\cite{FernandezNavarro:2023rhv,Davighi:2023evx,Barbieri:2023qpf} to address the flavour hierarchies. Our extension, based on Higgs compositeness, also provides a natural solution to the electroweak hierarchy problem, leaving only a minimised (and unavoidable) little hierarchy. Our analysis is limited to studying the effective theory describing the pNGBs (and their couplings to fermions and gauge fields) below the scale $4\pi f$. In the future, it would be interesting to formulate a candidate UV completion for the global symmetry breaking pattern we proposed here.

In a given UV completion, it is moreover natural to envisage that the gauge group we consider is embedded  into a semi-simple structure, as proposed in Ref.~\cite{Davighi:2023iks}, leading to a richer flavour phenomenology. As we have shown, the presence of TeV-scale composite dynamics allows the scale at which this embedding occurs to be parametrically separated from the electroweak scale. In the limit of a large separation, we obtain a realistic framework that simultaneously addresses the Higgs and flavour hierarchies, featuring TeV dynamics that are maximally safe from precision flavour constraints and best probed by high-$p_T$ searches and electroweak precision measurements.

\bigskip

\section*{Acknowledgement}

We are grateful to Sebastiano Covone and Marko Pesut for ongoing collaboration on related projects, and to Riccardo Rattazzi for discussions.
This research was
supported by the Swiss National Science Foundation, projects No. PCEFP2-194272 and
2000-1-240011.

\bibliographystyle{JHEP}
\bibliography{references.bib}

\appendix

\section{Estimating the pNGB potential} \label{app:quads}

The global $Sp(6)$ symmetry, whose spontaneous breaking to $SU(2)^3$ generates the three pNGB multiplets~\eqref{eq:pNGBs}, is explicitly broken by gauging the subgroup~\eqref{eq:gauging}, and by couplings to fermions responsible for generating Yukawa couplings. Both sets of interactions produce one-loop contributions to the potential for the pNGBs, responsible for the pNGB masses/vevs~\eqref{eq:quads} and interactions {\em e.g.}~\eqref{eq:cubic}. 

\medskip

\paragraph{Fermion-pNGB couplings.}
As explained in the main text,
gauge-invariance and the coset construction fix the fermion bilinear Lagrangian up to form factors.
In addition to those written in~\eqref{eq:chiral_flip} that flip chirality and give rise to Yukawa couplings, 
there are also chirality-preserving form factors that dress the kinetic terms for the chiral fermions, {\em e.g.}
\begin{align} \label{eq:top-2pt}
    \L \supset \overline{q}_L \slashed{p} \bigg[\Pi_0^{q_L} + \Pi_{1,3}^{q_L}\sin^2\left(\frac{h_{3}}{f}\right)
    + \Pi_{1,12}^{q_L}\sin^2\left(\frac{h_{12}}{f}\right) \bigg]q_L 
\end{align}
for the LH quark fields, which couple to the $H_3$ and $H_{12}$ pNGB multiplets. All such form factors $\Pi$ and $Y$ (in~\eqref{eq:chiral_flip}) are functions of $p^2$. We have $\Pi_0^{q_L}(0)=1$ {\em etc}.
By taking the inverse of the sum of ~\eqref{eq:chiral_flip} and~\eqref{eq:top-2pt} we can extract the Goldstone-dressed propagators for the SM chiral fermions.

As described in the main  text, 
the dependence of form factors on $M_\ast$, $g_\ast$, and $\Lambda$ depends on the dynamical generation of the operator: motivated primarily by phenomenological considerations, we invoke PC for third-generation interactions with $H_3$, but direct bilinear couplings between the light generations and $H_{12}$.  
For the PC-like form factors, we expect
\begin{equation} \label{eq:Y3_app}
    Y_{i3}^{u,d} \sim \frac{1}{g_\ast} \lambda_i^L(\Lambda) \lambda_{t,b}^{R\ast}(\Lambda) \left(\frac{M_\ast}{\Lambda}\right)^{\gamma_L - \gamma_R}
\end{equation}
as per~\eqref{eq:Y3}.
For the direct bilinear, we expect a different scaling:
\begin{equation}
    Y_{ij} \sim \alpha_{ij} \left(\frac{M_\ast}{\Lambda}\right)^{\gamma_{LR}}, \quad \gamma_{LR} \gtrsim 1 
\end{equation}
in addition to the different (rank-2) flavour structure.

\medskip

\paragraph{Gauge field-pNGB couplings.}
Each pNGB multiplet is a bidoublet under a pair of global $SU(2)$ factors, of which some generators are gauged. 
This generates couplings between the gauge fields and the pNGBs. For instance, we can parametrise the dressed gauge kinetic term as
\begin{align}
    \L_{AA} = \sum_{A\in \mathrm{Lie}(K)}\frac{1}{2}\left(\eta^{\mu\nu}-\frac{q^{\mu} q^{\nu}}{q^2}\right) \Pi_{\mu\nu}^A
\end{align}    
in the unbroken $K$ phase, 
where for instance the contributions relevant to $H_3$ are
\begin{align*}    
    &\qquad \qquad \Pi_{\mu\nu}^A  =  \Pi_0^A\left(A_{L,\mu}^a A_{L,\nu}^a + A_{R,\mu}^3 A_{R,\nu}^3 \right) + \\ 
    & \Pi_{L,R_3}^A \sin^2\left(\frac{h_3}{f}\right)\left(g_{L}A_{L,\mu}^a-g_{R}A_{R,\mu}^a\right)  \left(g_{L}A_{L,\nu}^a-g_{R}A_{R,\nu}^a\right)
\end{align*}
The kinetic term fixes $\Pi_0^A = -q^2$.
The leading term in $\Pi_{L,R_3}^A$ comes from the kinetic term of $H_3$, {\em viz.} $\L \supset \frac{f^2}{8}\mathrm{Tr} |DU|^2$ where $U=\exp(i\sqrt{2}\phi^a X^a/f)$ is the usual pNGB field, with $\phi^a$ collectively denoting all the Goldstones and $X^a$ running over all broken generators; matching to this yields $\Pi_{L,R_3}^A(0)=f^2/4$.

\medskip

\paragraph{Form factor ansatze.}
Given these couplings, we want to estimate the one-loop generation of the pNGB potential \`a la Coleman--Weinberg (CW), paying particular attention to the quadatics and cubics that are most important for understanding the dynamics of our model.

In order to estimate the loop integrals that generate the pNGB potential, we need to make an ansatz for the form factors away from $q^2=0$. We take:
\begin{align} \label{eq:FF}
    Y(q^2) = \frac{Y(0) M_{T}^2}{M_{T}^2-q^2}, \quad  
    \Pi_{XY}^A(q^2)=\frac{\Pi_{XY}^A(0) M_{\rho}^2}{M_{\rho}^2-q^2}    
\end{align}
corresponding to the simple assumption that each form factor is dominated by a quasi-isolated resonance at low $q^2$.

\medskip

\paragraph{Quadratics.}

The loop integrals generating the quadratic terms in the pNGB potential follows {\em e.g.} the discussion in~\cite{Covone:2024elw}. 
The Higgs field $H_3$ inevitably receives both gauge and fermion contributions. The gauge contributions are regulated by~\eqref{eq:FF} and are necessarily positive, while the fermion loops remain log-divergent. To match the observed Higgs mass and vev, $|M_3^2| \sim \mathcal{O}(0.1f)$, we need the fermion contribution to be negative and to be fine-tuned against the gauge contribution.
One way to cut the log-divergence down to size is to invoke a maximal left-right symmetry in the PC couplings that generate $y_t$, as in Ref.~\cite{Csaki:2017cep}. This gives
\begin{equation}
    M_3^2 \approx \frac{1}{4\pi^2}\left(-N_c y_t^2 M_{T}^2 + \frac{9}{8}(g_L^2+g_{R_3}^2) M_{\rho}^2 \right) \nonumber
\end{equation}
as per~\eqref{eq:Hmass} in the main text.

In contrast, we seek a spectrum in which the mass squared for the link field $\Sigma$ is negative and of order $f$ {\em i.e.} not fine-tuned. One natural scenario is for there to be no protecting left-right symmetry in the fermion couplings to $\Sigma$, in which case $M_\Sigma^2$ is expected to be dominated by the divergent fermion contribution, 
\begin{equation}
    M_\Sigma^2 \approx \frac{N_c}{8\pi^2} \kappa_f (4 \pi f)^2, \qquad \kappa_f < 0\, .
\end{equation}
Note the natural cut-off here is $4\pi f$, rather than $M_T$ as appears in~\eqref{eq:Hmass}. The sign of $\kappa_f$ is a free parameter in the underlying model, and we suppose it is negative to generate the vev for $\Sigma$.

On the other hand, for $M_{12}$, we know the couplings to fermions go through the direct bilinears and so are naturally small, $|\hat{y}_{ij}| \lesssim 0.1$, which is tied to our explanation of the $y_2/y_3$ Yukawa hierarchies. This means that, for $H_{12}$, it is natural to expect the gauge contribution to $M_{12}^2$ to be the dominant one, and this is necessarily positive:
\begin{equation}
    M_{12}^2 \approx \frac{9}{32\pi^2} (g_L^2 + g_X^2) M_\rho^2
\end{equation}
Because this is largely uncancelled by the fermion loop, we expect this to generate a TeV scale mass for $H_{12}$, and no vev. Note that $M_\rho \sim 8$ TeV or so is demanded by phenomenology, meaning $M_{12} \sim $ TeV is expected.

\medskip

\paragraph{Cubic.}

The key new ingredient we need to consider in this model is the generation of cubic scalar interactions that involve three different pNGB multiplets. The $H_{3} \Sigma H_{12}$ interaction of interest is obtained from the leading term in the expansion of the logarithm appearing in the CW formula for the potential. 

\begin{figure}[t]
    \begin{tikzpicture}[baseline=-4.0ex]
    \begin{feynman}   
        \vertex (a);
        \node at (a) [circle,fill,inner sep=1.2pt]{};
        \vertex [below=0.6in of a] (d);
        \vertex [right=0.25in of d] (e);
        \node at (e) [circle,fill,inner sep=1.2pt]{};
        \vertex [left=0.25in of d] (f);
        \node at (f) [circle,fill,inner sep=1.2pt]{};
        \vertex [above=0.3in of a] (aa) {$H_3$};
        \vertex [above left=0.3in of a] (a2);
        \vertex [above right=0.3in of a] (a3);
        \vertex [above=0.3in of a] (a4);
        \vertex [right=0.3in of e] (g) {$\Sigma$};
        \vertex [above right=0.3in of e] (g2);
        \vertex [below right=0.3in of e] (g3);
        \vertex [right=0.3in of e] (g4);
        \vertex [left=0.3in of f] (h) {$H_{12}$};
        \vertex [above left=0.3in of f] (h2);
        \vertex [below left=0.3in of f] (h3);
        \vertex [left=0.3in of f] (h4);
        \vertex (aa);
        \diagram* {
        (a) -- [scalar] (a4), 
        (a) -- [quarter left,fermion, edge label=$t_R$] (e), 
        (a) -- [quarter right,anti fermion,edge label'=$t_L$] (f),
        (e) -- [left,thick,fermion,edge label'=$\chi_R$] (f), 
        (e) -- [scalar] (g4),
        (f) -- [scalar] (h4),
        };
    \end{feynman}
    \end{tikzpicture} 
\caption{\label{fig:3Lloop} One-loop diagram generating the 
scalar trilinear coupling.}
\end{figure}
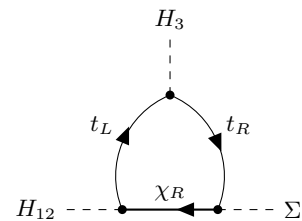

The generation of this coupling is, like other terms in the potential, model-dependent. To set out a concrete scenario, we suppose there is a vector-like fermion $(\chi_L, \chi_R)$ carrying global $U(1)_{B-L}^3$ charge, of mass $M_\chi$, which can run in the following one-loop diagram in 
Fig.~\ref{fig:3Lloop}.

The loop integral is
\begin{align}
    V_{H\Sigma H} =& -2N_c  \int \frac{d^4 p}{(2\pi)^4} 
    \, \hat{y}_{32} 
      f \sin\left(\frac{\sigma}{f}\right) 
     \frac{ \Pi_\chi (p) }{M_\chi}\,
    \times \nonumber \\
    & \frac{Y_{33}^u(p) f^2 \sin\left(\frac{h_3}{f}\right)  \sin\left(\frac{h_{12}}{f}\right)}{\slashed{p}(\Pi_0^{t_L}+\delta\Pi_{t_L}) \,\slashed{p}(\Pi_0^{t_R}+\delta\Pi_{t_R})
    }\,, 
\end{align}
where the $\delta \Pi$s in the denominator include the various pNGB contributions to the chiral fermion propagators, as in e.g.~\eqref{eq:top-2pt}. 
Notice the heavy fermion propagator is reduced to a contact interaction connecting the $H_{12}$ and $\Sigma$ insertions, with a $\Pi_\chi(p)/M_\chi$ suppression.
We have dropped subleading terms, including the cosine factors in the form factors~\eqref{eq:chiral_flip}.
The factor $2N_c$ comes from tracing over colours and spins, while the minus sign is due to the closed fermion loop.

Isolating the cubic scalar vertex from here, and taking the limit $M_\chi \gtrsim M_T$, we have
\begin{align}
    V_{H\Sigma H} \approx  &-8N_c \frac{\mathrm{Tr}[H_{3U} \Sigma_L H_{lU}^\dagger]}{f^3}  \Omega_4 \int\frac{p^3 dp}{(2\pi)^4} \\
    \times y_t \hat{y}_{32} \frac{f^3}{p^2}&\frac{ M_{T}^2}{(M_
    {T}^2+p^2)}  \frac{\Pi_\chi(0)}{M_\chi}
    + \mathcal{O}(\Phi^4) \nonumber
\end{align}
where we also Euclideanised. 
The loop integral converges in the UV.
We get
\begin{align}
    &V_{H\Sigma H} = -\frac{ N_c}{4\pi^2}
     y_t \hat{y}_{32} 
     \frac{M_T^2 \Pi_\chi(0)}{M_\chi}  \mathrm{Tr}[H_{3U} \Sigma_L H_{lU}^\dagger],
\end{align}   
as used in~\eqref{eq:cubic} in the main text.

\section{ FCNCs couplings of $H_{12}$} \label{app:H12}

The coupling of the $H_{12}$ to (light) fermion bilinears ($i,j=\{1,2\}$) is 
\begin{equation}
  {\mathcal L} \supset \frac{1}{\kappa}\, y_{ij}\, {\bar \psi}_{L}^{i}H_{12}\psi_R^j\,.
  \label{eq:yH12}
\end{equation}
To determine the couplings in the mass-eigenstate basis is useful to express $y_{ij}$  in terms of the full $3\times 3$ Yukawa coupling. 
Focusing for concreteness on the $q_L$-$d_R$
interaction we can write 
\begin{align}
 & {\mathcal L} \supset 
  \frac{1}{\kappa}\,  \overline{q}^i_L \left[ (1- {\hat P}_{3})  Y_d  (1- {\hat P}_{3})  \right]_{ij} d^i_R\,  H_{12}\,,
  \\
& ({\hat P}_{3})_{ij} = \delta_{i3}\delta_{j3}\,,
\qquad 
i,j=\{1,3\}\,.
  \label{eq:yPP}
\end{align}
When moving to the down-quark mass-eigenstate basis, the flavour coupling assumes the form 
\begin{equation}
\Delta =  (U_L^d)^\dagger \left[ (1- {\hat P}_{3})  Y_d  (1- {\hat P}_{3})  \right] U_R^d
\end{equation}
where, by construction, 
\begin{equation}
  (U_L^d)^\dagger  Y_d  U_R^d ={\rm diag}(y_d,y_s,y_b) \equiv \lambda_d\,.
\end{equation}
Given the general form of the Yukawa couplings 
in Eq.~\eqref{eq:yuk}, right-handed rotations are negligible and
\begin{equation}
\Delta \approx  [\lambda_d - \lambda_d {\hat P}_{3}]
- [ (U_L^d)^\dagger {\hat P}_{3} U_L^d \lambda_d
-   (U_L^d)^\dagger {\hat P}_{3} U_L^d \lambda_d {\hat P}_{3} ]\,.
\nonumber 
\end{equation}
This implies that the only effective FCNC
coupling of $H_{12}$ involves two light quarks 
and assumes the form 
\begin{equation}
  {\mathcal L} \supset   C^H_{ij} \bar d^i_L   d^i_R\,  H_{12}\,,
  \label{eq:yPP}
\end{equation}
with 
\begin{equation}
   C^H_{ij} = \frac{1}{\kappa} [ (U_L^d)^\dagger {\hat P}_{3}   U_L^d   \lambda_d ]_{ij}
   \sim \frac{1}{\kappa} V^*_{i3} V_{3j} y_{d_j}\,.
\end{equation}
Here the r.h.s.~is the maximal size expected under  up-alignment for the for left-handed quark doublets.
The double CKM and Yukawa suppression, as in MFV, warrant a sufficient protection despite the 
$\kappa^{-1}$ enhancement. 

\section{\boldmath{$Z^\prime$} couplings to SM}
\label{app:Zprime}

Near the TeV we have the gauge breaking
\begin{equation}
    \langle \Sigma \rangle = \frac{v_\Sigma}{\sqrt{2}} : U(1)_X \times U(1)_{R,3} \to U(1)_Y\, ,
\end{equation}
where $X=Y_{12}+(B-L)_3$.
This gives a massless hypercharge gauge boson $B_\mu \sim \frac{R_\mu}{g_R}+\frac{X_\mu}{g_X}$ and a massive $Z^\prime \sim \frac{R_\mu}{g_X}-\frac{X_\mu}{g_R}$, with mass $M_{R} =\frac{v_\Sigma}{2}\sqrt{g_R^2+g_X^2}$. The hypercharge gauge coupling of the unbroken $B_\mu$ field satisfies the matching condition $g_Y = g_X \cos\theta=g_R \sin\theta$, which defines the gauge mixing angle $\theta$. The `third family alignment' limit corresponds to $g_R \gg g_X$, so $\theta \to 0^+$.

The $Z^\prime$ couples to the SM current
\begin{equation} \label{eq:Zp_current}
    \mathcal{L} \supset  g_Y Z^\prime_\mu\left[ \cot\theta J_{R_3}^\mu - \tan\theta (J_{Y_{12}}^\mu+J_{B_3-L_3}^\mu) \right]
\end{equation}
The fermionic pieces of these currents are obvious, while the non-zero scalar part is $J_{R_3}^\mu \supset \frac{i}{2} H^\dagger D_\mu H + \mathrm{h.c.}$

\medskip

\paragraph{Flavour violation.}

After rotating to the fermion mass basis, one expects a flavour non-universal $Z^\prime$ such as ours to give tree-level (but CKM-suppressed) flavour violation.

However, due to the carefully crafted flavour structure of our model, it conspires that there is no flavour violation at leading order. This is because the TeV scale gauge breaking pattern is actually flavour universal in LH fields, while the asymmetric form of the Yukawa matrices~\eqref{eq:yuk} renders the RH mixing to be doubly-suppressed, as we used in App.~\ref{app:H12}.

Let us explain how this works in a little more detail. Given we have negligible RH fermion mixing, consider couplings of the $Z^\prime$ to LH quarks:
\begin{equation}
    \mathcal{L} \supset \tan\theta  g_Y Z^\prime \bar{f}_L^i [-(1-\hat{P}_3)\mathsf{y}_f - \hat{P}_3 ( \mathsf{b-l})_f ] f_L^j
\end{equation}
Going to the mass basis, the flavour structure (inside the square brackets) becomes
\begin{align}
    &U_{L}^{f\dagger} \left[-(1-\hat{P}_3)\mathsf{y}_f - \hat{P}_3 ( \mathsf{b-l})_f  \right] U_{L}^{f} \\
    \xrightarrow{\text{FV only}}\,\, &U_{L}^{f\dagger} \hat{P}_3 U_{L}^{f}\, \left(\mathsf{y}_f - ( \mathsf{b-l})_f \right) 
\end{align}
But for LH fields, $\mathsf{y}_f - ( \mathsf{b-l})_f = 0$. So, we conclude, there is no flavour violation at all, up to the doubly-suppressed RH mixing. 

We thus expect the $Z^\prime$ contributions to observables such as $\mathcal{B}(B_s \to \mu^+\mu^-)$ or $B_s-\overline{B}_s$ mixing parameters to give very weak constraints compared to those coming from flavour-conserving effects, which we turn to next. This is to be contrasted with other flavour deconstruction models, for which $B_s\to \mu^+\mu^-$ provides a leading probe~\cite{Davighi:2023evx,Davighi:2023xqn}.

\medskip

\paragraph{Electroweak.} Assume $M_{R}\gg m_Z$, and integrate out the $Z^\prime$. We get 
\begin{equation}
    C_{HD} = -\frac{g_Y^2 \cot^2\theta}{2 M_R^2}
\end{equation}
which results in a positive shift to the $W$ mass. We also modify the $Z$ boson couplings, via the following SMEFT operators:
\begin{align*}
    C_{Hu}^{33} &= \frac{g_Y^2 }{12M_R^2}\left(1-3\cot^2\theta\right)  , \quad  
    C_{Hu}^{ii} =  \frac{g_Y^2 }{3M_R^2} , \\
    C_{Hd}^{33} &= \frac{g_Y^2 }{12M_R^2}\left(1+3\cot^2\theta\right) ,  \quad  
    C_{Hd}^{ii} =  -\frac{g_Y^2 }{6M_R^2} ,  \\   
    C_{He}^{33} &= \frac{g_Y^2 }{4M_R^2}\left(-1+\cot^2\theta \right)  , \quad  
    C_{He}^{ii} =  -\frac{g_Y^2 }{2M_R^2} , \\
    C_{Hq}^{(1),33} &= \frac{g_Y^2 }{12M_R^2} = C_{Hq}^{(1),ii},   \quad  
    C_{Hl}^{(1),33} = -\frac{g_Y^2 }{4M_R^2}    = C_{Hl}^{(1),ii}
\end{align*}
where $i=1,2$ labels a light fermion index.
Note that the contributions from the $U(1)_X$ part (that acts on the light generations) is independent of the gauge mixing angle $\theta$, because they involve one third family field (the Higgs) and one light family field (the fermion current). Note also that the operators involving left-handed fermions are flavour-universal.

\medskip

\paragraph{High $p_T$.}
We also generate a slew of dimension-6 semi-leptonic operators. The flavour-conserving parts are order-1 in units of $g_Y^2/M_R^2$, and contribute to high-$p_T$ observables like $pp\to\ell\ell$ cross-section measurements in ATLAS and CMS. 
From~\eqref{eq:Zp_current}, the relevant operators generated can be identified by taking the 4-fermion pieces from:
\begin{equation}
    \mathcal{L} \supset -\frac{g_Y^2}{M_R^2} \left[ \cot\theta J_{R_3}^\mu - \tan\theta (J_{Y_{12}}^\mu+J_{B_3-L_3}^\mu) \right]^2
\end{equation}
For example, the LH-only operators are generated with the following (flavour-universal) Wilson coefficients, 
\begin{equation}
     C_{lq}^{(1)\alpha\alpha\beta\beta} =\frac{g_Y^2}{12 M_R^2} \tan^2\theta, \qquad \alpha, \beta = 1, 2, 3
\end{equation}
We do not write out all relevant WCs for brevity, but refer the reader to the matching done in Ref.~\cite{Davighi:2023evx} for a very similar model. Of course, all relevant contributions are included in our computation of the bounds from LHC Drell--Yan measurements using the \texttt{HighPT} package~\cite{Allwicher:2022mcg}, as described in the main text.

\medskip

\section{Absence of WZW term}

For a theory of pNGBs on some $G/H$, it is possible that topological interactions play an important role in the phenomenology. In our case, with $G/H=Sp(6)/SU(2)^3$, it is straightforward to show there is no Wess--Zumino--Witten (WZW) term. 
For our purposes here, we consider a WZW term~\cite{Wess:1971yu,Witten:1983tw} to be characterised by a non-trivial class in the free part of $H^5(G/H; \Z)$~\cite{DHoker:1994rdl}, corresponding to an integer-quantised topological term typically fixed by anomaly matching.

To see that there is no such term, we just need to show that $H^5(G/H)=0$. 
Recall that $Sp(6)$ can, as a topological space, be realised as a fibration $S^3 \hookrightarrow Sp(6) \to B$
where the base space $B$ is itself a fibration
$S^7 \hookrightarrow B \to S^{11}$.
The homology ring is, accordingly, $H_k(Sp(6)) = \Z[x_3, x_7, x_{11}]$, where $x_i$ has degree $i$.
The unbroken subgroup $H=SU(2)^3$ is a product of three 3-spheres, with $H_k(SU(2)^3)=\Z[y_3^{(1)}, y_3^{(2)},y_3^{(3)}]$, with $y_3^{(i)}$ the fundamental class of the $i^{\text{th}}$ $SU(2)$ factor.
From here, the homology of $X:=G/H$ follows from the long exact sequence (LES) in homology applied to the fibration
    $SU(2)^3 \hookrightarrow Sp(6) \to X$,
where the first map is the subgroup embedding of $H$ in $G$.
There is a segment
\begin{equation}
    \dots \to H_5(Sp(6)) \to H_5(X) \to H_4(SU(2)^3) \to \dots
\end{equation}
that reads
$\dots \to 0 \to H_5(X) \to 0 \to \dots$
Exactness implies 
    $H_5(X) = 0$,
hence the free part of the cohomology group $H^5(X)$ also vanishes,
so there is no quantized WZW term in this theory.

Even though we do not venture a UV completion of the $Sp(6) \to SU(2)^3$ symmetry breaking pattern in this work, the absence of the WZW term gives a clue, since it suggests the underlying theory has no chiral anomaly, hinting at UV dynamics with real or pseudo-real fermions.

Finally, we note that even though this simple topological argument precludes the existence of `cohomologically non-trivial' WZW terms, such as those that match anomalies or higher-group structures~\cite{Davighi:2024zjp}, there could still be non-zero topological interactions permitted in the effective action that have $\R$-valued coefficients. Examples like this appear in a composite Higgs model on the coset $SO(6)/SO(4)$, for instance~\cite{Davighi:2018xwn}.

\end{document}